\documentclass[preprint,preprintnumbers]{revtex4}

\usepackage{graphicx}
\usepackage{dcolumn}
\usepackage{bm}
\usepackage{epsfig}
\usepackage{epstopdf}
\usepackage{grffile}
\usepackage{color}
\usepackage{colordvi}
\usepackage{array}
\usepackage{tikz} 
\usepackage{tikz-feynman}
\usepackage{amsmath}
\usepackage{amssymb}
\usepackage{rotating}
\usepackage{lscape}
\usepackage{float}
\usepackage{footnote}
\usepackage{wasysym}
\usepackage{gensymb}
\usepackage{subcaption}
 \makesavenoteenv{environmentname}

\usepackage{hyperref}
\usepackage{ragged2e}
\makeatletter
\long\def\@makecaption#1#2{%
	\vskip\abovecaptionskip
	\sbox\@tempboxa{#1: #2}%
	\ifdim \wd\@tempboxa >\hsize
	#1: \justifying #2\par
	\else
	\global \@minipagefalse
	\hb@xt@\hsize{\hfil\box\@tempboxa\hfil}%
	\fi
	\vskip\belowcaptionskip}
\makeatother
\begin{document}


\title{Sterile Neutrino as an Asymmetric Dark Matter}

\author{S. Peyman Zakeri}
\affiliation{Faculty of Physics, Yazd University, P.~O.~Box 89195--741, Yazd, Iran}

\date{\today}

\begin{abstract}
We propose a minimal and predictive framework for asymmetric sterile neutrino dark matter (DM) produced via freeze-in. The standard model (SM) is extended by a gauge-singlet Dirac sterile neutrino carrying a conserved dark charge, a real scalar mediator, and an auxiliary singlet fermion. DM is generated through the out-of-equilibrium decay of the mediator, which simultaneously produces a particle–antiparticle asymmetry in the sterile sector controlled by a CP-violating parameter. We show that the observed relic abundance can be naturally reproduced without thermal equilibration with the SM plasma. The resulting non-thermal momentum distribution is colder than a thermal Fermi--Dirac spectrum, ensuring consistency with structure formation constraints. Combining relic density, Lyman-$\alpha$, Higgs invisible decay, and big bang nucleosynthesis (BBN) bounds, we identify correlated and predictive regions of the parameter space characterized by non-trivial relations among the sterile neutrino mass and the decay parameters. This scenario provides a self-consistent realization of Dirac asymmetric sterile neutrino DM within an asymmetric freeze-in (AFI) framework, offering a constrained and testable alternative to conventional production mechanisms.
\end{abstract}

\maketitle


\section{Introduction}\label{sec:int} 
A wide range of astrophysical and cosmological observations indicate that our Universe is dominated by a non-luminous component of matter, commonly referred to as DM, whose particle nature remains unknown~\cite{Bertone:2005,Feng:2010}. In the standard cosmological paradigm, DM is required to be stable on cosmological timescales, electrically neutral, and sufficiently cold to seed bottom-up large-scale structure formation.

Traditionally, the origins of DM and baryons have been treated as independent phenomena. In the conventional weakly interacting massive particle (WIMP) paradigm, the DM relic abundance is set by thermal freeze-out~\cite{Zeldovich:1965,Chiu:1966}, resulting in a symmetric population of particles and antiparticles with no conserved quantum number. In contrast, the baryon abundance arises from a particle-antiparticle asymmetry generated through CP-violating and baryon-number-violating interactions occurring out of equilibrium. Remarkably, observations indicate that the present-day energy densities of DM and baryons are of the same order, $\rho_{\mathrm{DM}} \simeq 5\rho_{\mathrm{B}}$~\cite{PDG,Petraki:2013, Zurek:2013}. This numerical proximity has motivated scenarios in which both abundances share a common origin. A simple way to relate the two sectors is to assign a conserved quantum number (such as $B-L$) to DM, leading to the framework of asymmetric DM (ADM)~\cite{Kaplan:2009}.

An alternative to freeze-out production is the freeze-in mechanism~\cite{Hall:2009}, in which DM particles are produced via extremely feeble interactions and never reach thermal equilibrium with the SM bath. In this scenario, the dark sector interacts so weakly with the visible sector that its abundance gradually builds up from rare decay or scattering processes. The AFI framework~\cite{Hall:2010} combines the virtues of ADM and freeze-in: the DM abundance is generated out of equilibrium and can simultaneously carry a particle-antiparticle asymmetry. Since freeze-in production is typically infrared dominated, it is largely insensitive to the unknown high-temperature history of the Universe. However, it has been shown that the efficiency of asymmetry generation can depend sensitively on the thermal history of the mediator sector. In particular, if the mediator responsible for transferring the asymmetry is in thermal equilibrium, the resulting asymmetry may be significantly suppressed~\cite{Hook:2011,Unwin:2014}.

Sterile neutrinos provide a particularly well-motivated DM candidate. The observation of neutrino oscillations already implies the existence of physics beyond the SM (BSM)~\cite{Minkowski:1977,GellMann:1979,Mohapatra:1979,Yanagida:1979}, and gauge-singlet fermions naturally arise in many extensions. Sterile neutrinos with keV-scale masses and tiny mixing with active neutrinos have long been considered viable DM candidates~\cite{Dodelson:1993}. In conventional scenarios, they are produced non-thermally via active-sterile oscillations~\cite{Barbieri:1990,Enqvist:1990}. However, this mechanism is strongly constrained by structure formation and X-ray observations~\cite{Canetti:2012,Canetti:2012b,Merle:2015}. Resonant production in the presence of a large primordial lepton asymmetry (the Shi--Fuller mechanism~\cite{Shi:1998}) can alleviate some of these bounds, although the viable parameter space remains restricted.

Motivated by the AFI framework and the possibility of generating a sterile-sector asymmetry through feeble interactions, we explore a scenario in which a naturally light sterile neutrino is produced asymmetrically through freeze-in decays of a scalar mediator. In contrast to the extensively studied symmetric production of sterile neutrinos from scalar decays~\cite{Merle:2013,Merle:2015b, Konig:2016, Drewes:2015, Petraki:2007}, we focus on the generation of a sterile-sector particle-antiparticle asymmetry. Since a Majorana fermion is self-conjugate and cannot support a conserved dark charge, it cannot serve as an ADM candidate. We therefore consider a Dirac sterile neutrino carrying a conserved global symmetry, with its mass generated via a Higgs-type mechanism. This setup naturally accommodates asymmetric production within a consistent field-theoretic framework.

The paper is organized as follows. In Sec.~\ref{sec:model}, we present the model and its field content. In Sec.~\ref{sec:boltz}, we derive and solve the Boltzmann equations governing the asymmetric sterile neutrino abundance. The phenomenological implications and viable parameter space are discussed in Sec.~\ref{sec:aspect}. We conclude in Sec.~\ref{sec:conclud}.

\section{The Model}\label{sec:model}
The SM is extended by three gauge-singlet fermions $N_I$ $(I=1,2,3)$, a real scalar mediator $\phi$, and an auxiliary fermion $\chi$, all of which are singlets under the SM gauge group. The fields $N_I$ represent sterile (right-handed) neutrinos. From a theoretical perspective, there is no fundamental restriction on the number of sterile neutrinos, while experimental constraints arise mainly from their mixing with active neutrinos. For instance, two sterile neutrinos are sufficient to explain the observed neutrino mass splittings from solar and atmospheric oscillation data~\cite{Minkowski:1977,GellMann:1979,Mohapatra:1979,Yanagida:1979}, whereas additional sterile states may be motivated by phenomena such as DM, pulsar kicks, or cosmological considerations~\cite{Kusenko:2005,Kusenko:2005b}.

In this work, we consider a minimal realization in which only one sterile neutrino participates in the DM dynamics. We therefore focus on a single sterile neutrino, denoted by $N$, which plays the role of ADM. The remaining sterile states are assumed to be irrelevant for the phenomenology discussed here.

The model features a global $U(1)_{\rm DM}$ symmetry under which the sterile neutrino $N$ carries a nonzero dark charge, while its antiparticle $\bar N$ carries the opposite charge. This symmetry plays a role analogous to baryon number in the visible sector and ensures both the cosmological stability of $N$ and the preservation of the ADM abundance. Since all interactions of the model respect $U(1)_{\rm DM}$, any asymmetry generated between $N$ and $\bar N$ is conserved throughout the cosmological evolution, following the general framework of ADM scenarios~\cite{Unwin:2014,Falkowski:2011}. The auxiliary fermion $\chi$ is assigned the same dark charge, while the scalar mediator $\phi$ is taken to be neutral under this symmetry. 

The relevant interactions of the model are described by the following Lagrangian:
\begin{equation}
	\begin{aligned}
		\mathcal{L} \supset \;&
		\frac{1}{2}\partial_\mu \phi\, \partial^\mu \phi
		+ i \bar N \gamma^\mu \partial_\mu N
		+ i \bar \chi \gamma^\mu \partial_\mu \chi
		\\
		&
		- m_N \bar N N
		- m_\chi \bar \chi \chi
		\\
		&
		- \left(
		\lambda_\alpha \bar L_\alpha \tilde H N
		+ \mu\, \phi\, \bar \chi N
		+ \text{h.c.}
		\right)
		- V(H,\phi) ,
	\end{aligned}
	\label{eq:Lagrangian}
\end{equation}
where $H$ is the SM Higgs doublet, $\tilde H \equiv i\sigma_2 H^*$, and $L_\alpha = (\nu_{\alpha L}, \ell_\alpha)_L$ $(\alpha = e,\mu,\tau)$ denotes the SM lepton doublets. The Yukawa interaction involving $\phi$, $N$, and $\chi$ is taken to be chiral, ensuring that the dominant decay channel of the mediator is $\phi \rightarrow N + \chi$. In this regard, $N$ and $\chi$ are treated as Dirac fermions and the chiral projectors are suppressed in the notation. Additionally, the auxiliary fermion $\chi$ is assumed to be unstable due to additional feeble interactions not included in the minimal setup. These interactions are sufficiently weak not to affect the freeze-in production of DM, while allowing $\chi$ to decay before contributing appreciably to the cosmological relic density.

The scalar potential is given by
\begin{equation}
	V(H,\phi) =
	- \mu_H^2 H^\dagger H
	+ \lambda_H (H^\dagger H)^2
	+ \frac{1}{2} \mu_\phi^2 \phi^2
	+ \frac{\lambda_\phi}{4} \phi^4
	+ \lambda_{\phi H} \phi^2 H^\dagger H .
	\label{eq:potential}
\end{equation}
Linear and cubic terms in $\phi$ are forbidden by the imposed symmetries. The Higgs portal coupling $\lambda_{\phi H}$ governs the interaction between the visible and dark sectors and is responsible for maintaining thermal contact between the mediator and the SM plasma.

Therefore, the DM production mechanism is primarily controlled by the parameters
\[
m_N, \quad m_\phi, \quad \mu,
\]
together with the CP-violating parameter $\varepsilon$ associated with the decay of $\phi$. While the model also contains the auxiliary fermion mass $m_\chi$ and the Higgs-portal coupling $\lambda_{\phi H}$, these parameters have little impact on the phenomenological results discussed in this work.

\section{Relic abundance}\label{sec:boltz} 

\subsection{Boltzmann equation}\label{sec3.1}
The relic abundance of sterile neutrino DM is generated through the freeze-in production mechanism from the decays of the scalar mediator $\phi$. The mediator is assumed to remain in thermal equilibrium with the visible-sector plasma through the Higgs portal interaction, while the sterile fermion $N$ and the auxiliary fermion $\chi$ never thermalize owing to the feeble coupling $\mu$. Consequently, the DM asymmetry is generated through CP-violating decays of the mediator while the sterile sector itself remains out of thermal equilibrium. Since the sterile states are populated only through freeze-in, their initial abundance is negligible and inverse processes such as $\bar{\chi}N \rightarrow \phi$ can be safely neglected throughout the cosmological evolution~\cite{Hall:2009}.

The generation of the DM asymmetry is parametrized by the CP asymmetry parameter $\varepsilon$, which arises from the interference between tree-level and loop-level decay amplitudes of $\phi \rightarrow \bar{\chi} N$~\cite{Hook:2011,Unwin:2014}:
\begin{equation}
	\varepsilon = \frac{\Gamma(\phi \rightarrow \bar{\chi} N) - \Gamma(\phi \rightarrow \chi \bar{N})}{\Gamma_\phi},
\end{equation}
where $\Gamma_\phi$ is the total decay width of the scalar mediator.

The asymmetry yield is then defined as
\begin{equation}
	Y_- \equiv Y_N - Y_{\bar N}
	=
	\frac{n_N - n_{\bar N}}{s} \, ,
	\label{eq:yield}
\end{equation}
where $s$ denotes the entropy density of the Universe. Assuming that the scalar mediator remains in thermal equilibrium throughout the production epoch, such that $n_\phi \simeq n_\phi^{\rm eq}$, the evolution of the asymmetric number density $n_- = n_N - n_{\bar N}$ is described by the Boltzmann equation
\begin{equation}
	\frac{dn_-}{dt} + 3H n_- = \varepsilon \, \Gamma_{\phi} \, n_\phi^{\rm eq} \, .
	\label{eq:boltz}
\end{equation}
Here, $n_\phi^{\rm eq}$ is the equilibrium number density of the mediator~\cite{Gondolo:1991,Kolb:1990}.
Using the relation $d/dt = - H T \, d/dT$ and the entropy density
\begin{equation}
	s(T) = \frac{2\pi^2}{45} g_* T^3 \, ,
	\label{eq:entropy}
\end{equation}
with $g_*$ denoting the effective number of relativistic degrees of freedom, the Boltzmann equation can be expressed in terms of the yield as
\begin{equation}
	\frac{dY_-}{dT}
	=
	- \frac{\varepsilon \, \Gamma_\phi \, n_\phi^{\rm eq}(T)}
	{s(T)\, H(T)\, T} \, .
	\label{eq:boltz_yield}
\end{equation}
The equilibrium number density of the scalar mediator is given by
\begin{equation}
	n_\phi^{\rm eq}(T)
	=
	g_\phi\frac{m_\phi^2 T}{2\pi^2}
	K_2\!\left( \frac{m_\phi}{T} \right) \, ,
	\label{eq:neq}
\end{equation}
where $g_\phi$ is the number of internal degrees of freedom of the scalar mediator. Since $\phi$ is a real scalar field, we take $g_\phi = 1$ throughout this work. Furthermore, the Hubble expansion rate during radiation domination reads
\begin{equation}
	H(T) = 1.66 \, \sqrt{g_*} \, \frac{T^2}{M_{\rm Pl}} \, ,
	\label{eq:hubble}
\end{equation}
where $M_{\rm Pl}$ is the Planck mass. Finally, the initial condition is chosen as
\begin{equation}
	Y_-(T \gg m_\phi) = 0 \, ,
	\label{eq:initial}
\end{equation}
reflecting the absence of sterile neutrinos at early times.

Integrating the Boltzmann equation from the onset of production to the present epoch, the final asymmetry yield is obtained as
\begin{equation}
	Y_-^\infty
	=
	\varepsilon
	\int_0^\infty
	\frac{\Gamma_\phi \, n_\phi^{\rm eq}(T)}
	{s(T)\, H(T)\, T}
	\, dT \, .
	\label{eq:integral}
\end{equation}
By introducing the dimensionless variable $x = m_\phi/T$, the above integral can be evaluated analytically, yielding
\begin{equation}
	Y_-^\infty
	\simeq
	\frac{45}{1.66\, 4\pi^4}
	\frac{g_\phi}{g_*^{3/2}}
	\frac{M_{\rm Pl}}{m_\phi}
	\varepsilon \, \Gamma_\phi
	\int_0^\infty x^3 K_2(x)\, dx \, .
	\label{eq:integral_x}
\end{equation}
Using $\int_0^\infty x^3 K_2(x)\, dx = 8$, the final expression simplifies to
\begin{equation}
	Y_-^\infty
	\simeq
	\frac{90}{1.66\, \pi^4}
	\frac{g_\phi}{g_*^{3/2}}
	\frac{M_{\rm Pl}}{m_\phi^2}\varepsilon \, \Gamma_\phi \, .
	\label{eq:final_yield}
\end{equation}
Finally, the present-day DM relic density is related to the asymmetry yield by~\cite{Petraki:2013,Zurek:2013}
\begin{equation}
	\Omega_{\rm DM} h^2
	=
	2.75 \times 10^8
	\left( \frac{m_N}{\text{GeV}} \right)
	Y_-^\infty \, .
	\label{eq:relic}
\end{equation}
The relic abundance is therefore determined by the asymmetry yield generated through CP-violating mediator decays, confirming that the production mechanism proceeds in the freeze-in regime.

\subsection{The DM abundance}\label{sec3.2}
In this subsection, we present the numerical results obtained from solving the Boltzmann equations governing the production of asymmetric sterile neutrino DM. We analyze the thermal evolution of the sterile neutrino yield and investigate its dependence on the relevant model parameters, such as the interaction strength $\mu$, the CP-violating parameter $\varepsilon$, and the mediator mass $m_\phi$. The production mechanism is dominated by the freeze-in decay $\phi \rightarrow \bar{\chi} N$,  while the sterile sector remains out of thermal equilibrium. The final relic abundance is established once the production process effectively ceases.

\begin{figure}[!htbp]
	\centering
	\includegraphics[width=0.67\textwidth]{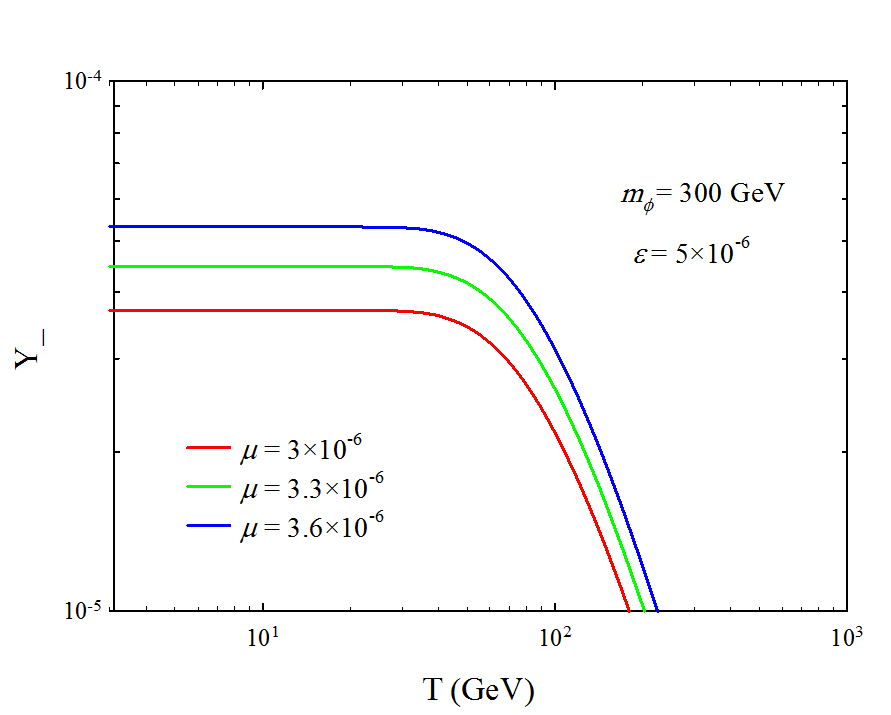}
	\caption{Evolution of the asymmetric sterile neutrino abundance $Y_{-}$ as a function of temperature for different values of the effective coupling $\mu$. The CP asymmetry parameter $\varepsilon$ and mediator mass $m_\phi$ are kept fixed.}
	\label{fig:Y_vs_T_mu}
\end{figure}

Figure~\ref{fig:Y_vs_T_mu} shows the evolution of the comoving asymmetric abundance $Y_{-}$ as a function of temperature for three different values of the effective coupling $\mu$, while $m_\phi$ and $\varepsilon$ are kept fixed. As expected, the production of the sterile neutrino asymmetry occurs predominantly at temperatures around $T \sim m_\phi$, where the mediator population is still abundant and CP-violating decays efficiently generate the asymmetry. Increasing $\mu$ enhances the decay rate of $\phi$ and leads to a larger final asymmetric yield, while the overall freeze-in nature of the production mechanism is preserved.

\begin{figure}[!htbp]
	\centering
	\includegraphics[width=0.67\textwidth]{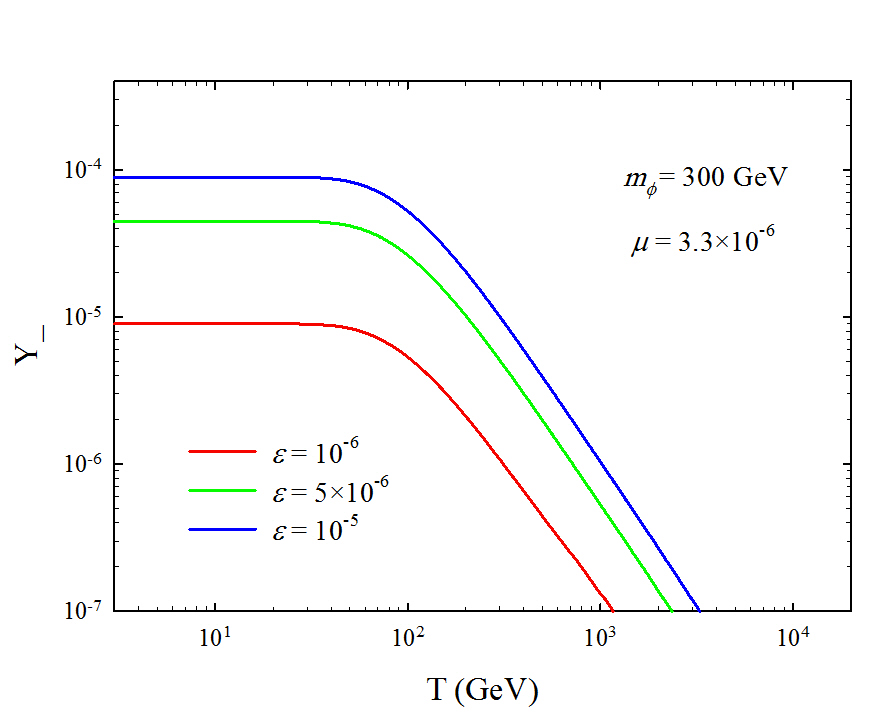}
	\caption{Evolution of the asymmetric sterile neutrino abundance $Y_{-}$ as a function of temperature for different values of the CP asymmetry parameter $\varepsilon$. The effective coupling $\mu$ and mediator mass $m_\phi$ are kept fixed.}
	\label{fig:Y_vs_T_eps}
\end{figure}

In Fig.~\ref{fig:Y_vs_T_eps}, we fix $m_\phi$ and $\mu$ and vary the CP asymmetry parameter $\varepsilon$. The final abundance scales approximately linearly with $\varepsilon$, reflecting the fact that the net sterile neutrino asymmetry is directly proportional to the CP-violating decay asymmetry of the mediator. This behavior confirms that the relic abundance is controlled primarily by the CP-violating source term responsible for generating the asymmetry.

\begin{figure}[!htbp]
	\centering
	\includegraphics[width=0.7\textwidth]{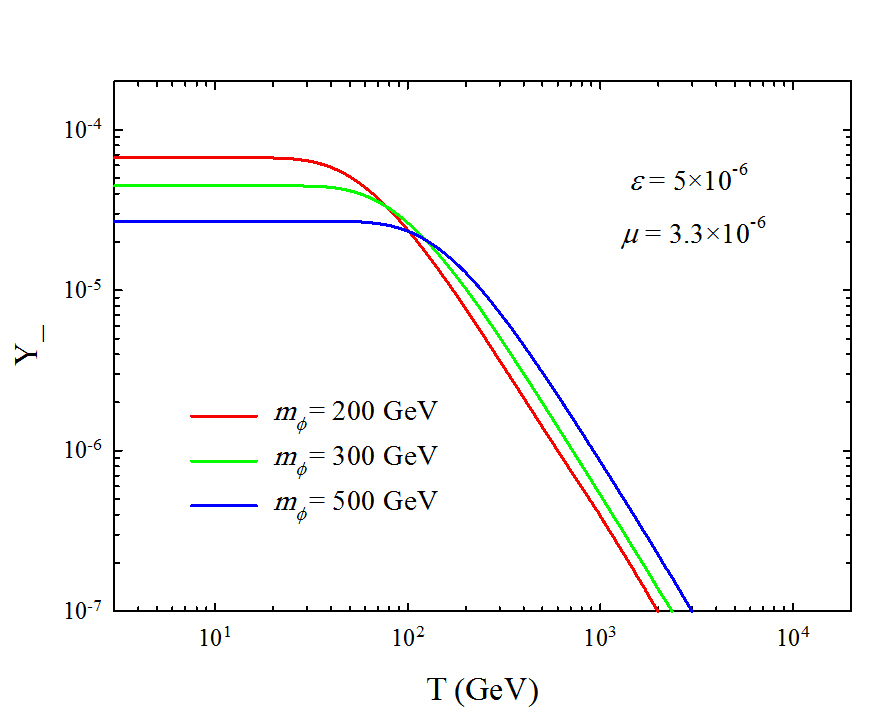}
	\caption{Evolution of the sterile neutrino yield $Y_{-}$ as a function of the temperature for different values of the mediator mass $m_\phi$. The other model parameters are fixed.
	}
	\label{fig:Y_vs_T_mphi}
\end{figure}

Figure~\ref{fig:Y_vs_T_mphi} shows the evolution of the sterile neutrino yield $Y_{-}$ as a function of the temperature for several benchmark values of the mediator mass $m_\phi$, while keeping the other parameters fixed. We observe that increasing the mediator mass shifts the production epoch to higher temperatures and slightly modifies the final frozen value of the yield. This behavior is expected, since the characteristic temperature of freeze-in production is set by the mediator mass, with the dominant contribution arising at $T \sim m_\phi$~\cite{Merle:2013,Merle:2015b}.

\begin{figure}[!htbp]
	\centering
	\includegraphics[width=0.7\textwidth]{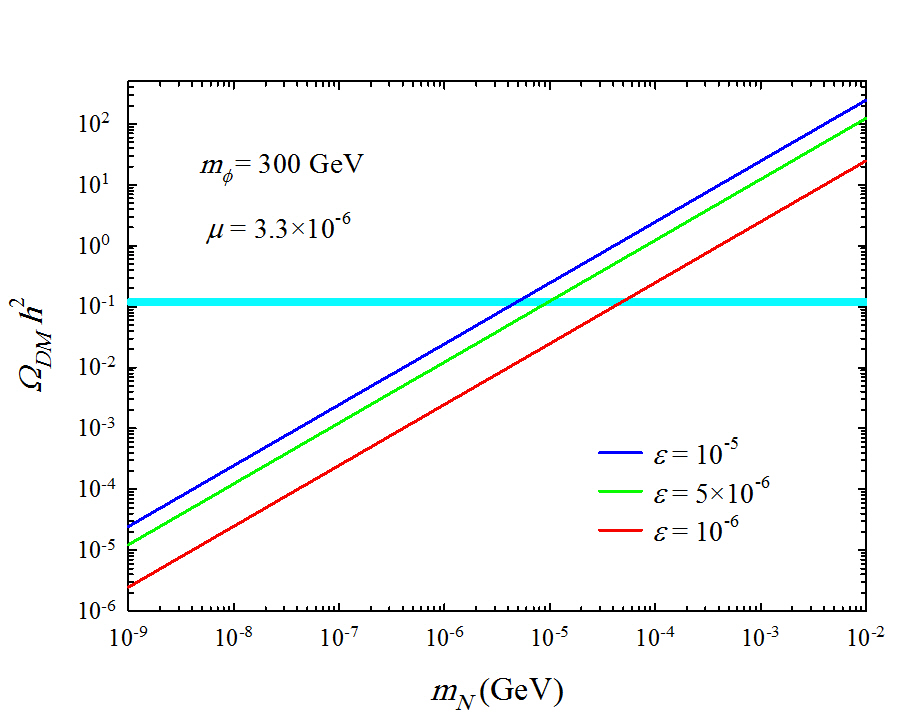}
	\caption{
		Relic density $\Omega_{\rm DM} h^2$ as a function of the sterile neutrino mass $m_N$.
		The other model parameters are kept fixed.
		The linear behavior reflects the proportionality
		$\Omega_{\rm DM} h^2 \propto m_N Y_-^\infty$
		characteristic of AFI production.
		The horizontal shaded band corresponds to the observed DM
		relic abundance, taken here as
	    $0.11 \le \Omega_{\rm DM} h^2 \le 0.13$,
		centered around the Planck value
		$\Omega_{\rm DM} h^2 \simeq 0.12$.
	}
	\label{fig:O_vs_mN}
\end{figure}

Finally, Fig.~\ref{fig:O_vs_mN} illustrates the relic density as a function of the sterile neutrino mass $m_N$ for different values of the asymmetry parameter $\varepsilon$. As expected from the relation $\Omega_{\rm DM} h^2 \propto m_N Y_-^\infty$, the relic density exhibits an approximately linear dependence on the DM mass. Since the freeze-in yield $Y_-^\infty$ is primarily determined by the mediator decay parameters and remains essentially independent of $m_N$, the scaling with $m_N$ directly reflects the mass contribution to the energy density. The observed DM relic abundance, $\Omega_{\rm DM} h^2 \simeq 0.12$~\cite{Planck:2018}, can therefore be reproduced for sterile neutrino masses within a specific range, demonstrating the viability of the AFI mechanism.

Overall, these numerical results demonstrate that the model naturally generates the correct DM abundance through AFI, with a clear and controlled dependence on the parameters $\mu$, $\varepsilon$, $m_{\phi}$ and $m_N$. The parameter space consistent with the observed relic density will be further constrained in Sec.~\ref{sec:aspect} by phenomenological considerations.

\section{Phenomenological aspects}\label{sec:aspect}
\subsection{Asymmetric sterile DM: cold or warm?} \label{sec4.1}
One of the crucial phenomenological questions in sterile neutrino DM scenarios concerns the nature of DM, namely whether it behaves as cold, warm, or hot DM. This classification is determined by the typical momentum distribution of DM particles at production, which directly affects structure formation at small scales~\cite{Blumenthal:1984,Primack:2000}. In AFI scenarios, the momentum spectrum of sterile neutrinos can significantly differ from that of thermal relics, making a dedicated analysis necessary~\cite{Merle:2013,Merle:2015b,Konig:2016}.

A convenient and widely used quantity to characterize the ``warmness'' of DM is the dimensionless average momentum-to-temperature ratio, defined as
\begin{equation}
	\left\langle \frac{p}{T} \right\rangle
	=
	\frac{\int d^3p \, \frac{p}{T} f_N(p,T)}
	{\int d^3p \, f_N(p,T)} \, .
	\label{eq:avg_p}
\end{equation}
Here, $f_N(p,T)$ denotes the phase-space distribution function of the sterile neutrino. In the present model, sterile neutrinos are dominantly produced through the decay of the scalar mediator $\phi$. Since the production occurs via freeze-in, the resulting momentum distribution of $N$ is non-thermal and sharply peaked at the kinematic value set by the two-body decay~\cite{Merle:2013,Merle:2015b}. To a good approximation, the corresponding momentum distribution can be represented as
\begin{equation}
	f_N(p,T) \propto 
	A(T)
	\delta\!\left(p - \frac{m_\phi}{2}\right) \, ,
	\label{eq:dist}
\end{equation}
where $A(T)$ denotes an overall normalization factor that does not affect the evaluation of the average momentum-to-temperature ratio.
Due to the two-body decay kinematics, sterile neutrinos are produced with a fixed momentum $p\simeq m_\phi/2$ at the time of production. This allows for a simple analytical estimate of the average momentum-to-temperature ratio, which can be evaluated at the characteristic production temperature $T_{\text{prod}}\sim m_\phi$~\cite{Merle:2013}
\begin{equation}
	\left\langle \frac{p}{T} \right\rangle
	\simeq \frac{m_\phi}{2T_{\rm prod}} \approx 0.5 \, .
	\label{eq:avg_p_est}
\end{equation}

The above estimate shows that the typical momentum of sterile neutrinos is suppressed compared to that of a relativistic thermal fermion, for which $\left\langle \frac{p}{T} \right\rangle\simeq 3.15$~\cite{Gondolo:1991}. Consequently, sterile neutrinos produced via AFI are expected to be colder than SM relics
\begin{equation}
	\left\langle \frac{p}{T} \right\rangle < 3.15.
	\label{eq:cold}
\end{equation}

Therefore, for the parameter space explored in this work, sterile neutrinos behave effectively as cold or mildly warm DM (WDM), consistent with large-scale structure formation and Lyman-$\alpha$ constraints~\cite{Viel:2005}. This feature represents an important phenomenological advantage of the AFI mechanism compared to conventional thermal production scenarios such as the Dodelson--Widrow mechanism~\cite{Dodelson:1993}.

\subsection{Higgs invisible decay} \label{sec4.2}
The Higgs-portal interaction provides a possible connection between the visible and dark sectors through the coupling $\lambda_{\phi H} \phi^2 H^\dagger H$. In principle, this interaction allows the invisible decay of the Higgs boson into a pair of scalar mediators,
\begin{equation}
	h \rightarrow \phi\phi ,
\end{equation}
provided that the kinematic condition $m_\phi < m_h/2$ is satisfied. Such invisible Higgs decays are constrained by collider measurements~\cite{ATLAS:2022,CMS:2023}.

In the parameter space considered in this work, however, the mediator mass lies in the range $m_\phi = 200\text{--}500~\mathrm{GeV}$,
which is well above the threshold $m_h/2 \simeq 62.5~\mathrm{GeV}$. Consequently, the decay channel $h \rightarrow \phi\phi$ is kinematically forbidden, and the Higgs invisible decay bound is automatically satisfied throughout the viable parameter space. Therefore, this constraint does not impose any additional restriction on the model parameters considered here.

\subsection{Lyman-$\alpha$} \label{sec4.3}
One of the most stringent constraints on WDM scenarios arises from Lyman-$\alpha$ forest observations, which probe the small-scale matter power spectrum at high redshifts~\cite{Croft:1998,McDonald:2005}. These measurements are particularly sensitive to the free-streaming length of DM particles and therefore to their momentum distribution~\cite{Viel:2005,Boyarsky:2009}.

In order to compare the sterile neutrino DM produced in our model with standard WDM bounds, it is convenient to introduce an effective WDM mass, defined by equating the free-streaming properties of the two distributions~\cite{Viel:2005}. This quantity is given by
\begin{equation}
	m_{\rm WDM}^{\rm eff}
	=
	m_N
	\left(
	\frac{\langle p/T \rangle_{\rm FD}}
	{\langle p/T \rangle}
	\right) \, .
	\label{eq:wdm_mass}
\end{equation}
Here, $\langle p/T \rangle$ denotes the average momentum-to-temperature ratio of sterile neutrinos produced via freeze-in decays, while $\langle p/T \rangle_{\rm FD}$ corresponds to the value obtained for a relativistic Fermi--Dirac distribution. For such a thermal spectrum, this average value is~\cite{Gondolo:1991}
\begin{equation}
	\langle p/T \rangle_{\rm FD} = 3.15 \, .
	\label{eq:fd_avg}
\end{equation}

Lyman-$\alpha$ observations impose a lower bound on the effective warm DM mass, which can be translated into a constraint on the sterile neutrino parameter space. A representative lower bound from recent Lyman-$\alpha$
analyses can be expressed as~\cite{Palanque-Delabrouille:2019,Irsic:2024}
\begin{equation}
	m_{\rm WDM}^{\rm eff} \gtrsim 5~{\rm keV} \, .
	\label{eq:lyman_bound}
\end{equation}
The precise numerical value depends on the adopted dataset
and assumptions regarding the thermal history of the intergalactic
medium. We therefore adopt $5$~keV as a representative reference value.

Since sterile neutrinos in our model are produced non-thermally and typically have a colder momentum distribution compared to a thermal relic, the corresponding Lyman-$\alpha$ bound is significantly relaxed. This allows a wide region of the parameter space, consistent
with the relic abundance and asymmetric production mechanism, to evade current Lyman-$\alpha$ constraints.

\subsection{BBN} \label{sec4.4}
BBN constrains any late-time energy injection or modification of the expansion rate of the Universe at temperatures $T \sim \mathcal{O}(\mathrm{MeV})$~\cite{Kawasaki:2018,Cyburt:2015,Pitrou:2018}. In the present model, sterile neutrinos are produced through the decay of the scalar field $\phi$. The decay takes place when the Hubble expansion rate satisfies
\begin{equation}
	\Gamma_\phi \simeq H(T_{\rm decay}) \, ,
	\label{eq:decay_condition}
\end{equation}
which leads to a decay temperature
\begin{equation}
T_{\rm decay}
\simeq
0.78\, g_*^{-1/4}
\sqrt{\Gamma_\phi M_{\rm Pl}} \, .
\label{eq:decay_temp}
\end{equation}

For the parameter space considered in this work, the decay temperature is well above the BBN scale, ensuring that the production of sterile neutrinos is completed before the onset of nucleosynthesis. As a result, no significant entropy injection or distortion of the neutron-to-proton ratio occurs during BBN~\cite{Kawasaki:2018}.

Moreover, since sterile neutrinos are produced non-thermally and remain decoupled from the SM plasma, their contribution to the radiation energy density is negligible at the BBN epoch. Consequently, the effective number of relativistic degrees of freedom remains essentially unchanged,
\begin{equation}
	\Delta N_{\rm eff} \ll 1 \, ,
	\label{eq:delta_neff}
\end{equation}
and the model safely satisfies all BBN constraints~\cite{Fields:2019,Pitrou:2018}.

\subsection{Viable parameter space} \label{sec4.5}
In this subsection, we determine the viable parameter space of the model by combining all relevant cosmological and phenomenological constraints. As stated before, the scalar mediator is assumed to remain in thermal equilibrium, while the sterile states are produced through feeble decays. In particular, we impose the requirement of reproducing the observed DM relic density, $\Omega_{\rm DM} h^2 \simeq 0.12$~\cite{Planck:2018}, together with the freeze-in production condition, $\Gamma_\phi < H(T \sim m_\phi)$~\cite{Hall:2009}, and the Lyman-$\alpha$ structure formation constraint~\cite{Palanque-Delabrouille:2019,Irsic:2024}.

As discussed in Sec.~\ref{sec:boltz}, the present-day relic density is given by
\begin{equation}
	\Omega_{\rm DM} h^2
	=
	2.75 \times 10^8
	\left(\frac{m_N}{\rm GeV}\right)
	Y_-^\infty ,
	\label{eq:relic_final}
\end{equation}
where $Y_-^\infty$ denotes the final asymmetric yield. The requirement of reproducing the observed DM abundance imposes a correlation among the model parameters, which is reflected in the diagonal bands shown in the following figures.

Figure~\ref{fig:mu_mN} shows the viable region in the $(m_N,\mu)$ plane for fixed values of $m_\phi = 300\,\mathrm{GeV}$ and $\varepsilon = 5 \times 10^{-6}$. The diagonal red band corresponds to the set of parameter points that simultaneously reproduce the observed relic density and satisfy the freeze-in requirement and other phenomenological bounds. The vertical dashed line indicates the lower bound on the sterile neutrino mass from Lyman-$\alpha$ forest observations, $m_N \gtrsim 5$~keV, derived in Sec.~\ref{sec4.3}. The physically viable region therefore lies in the overlap between the diagonal band and the region to the right of the Lyman-$\alpha$ bound.

\begin{figure}[!htbp]
	\centering
	\includegraphics[width=0.75\textwidth]{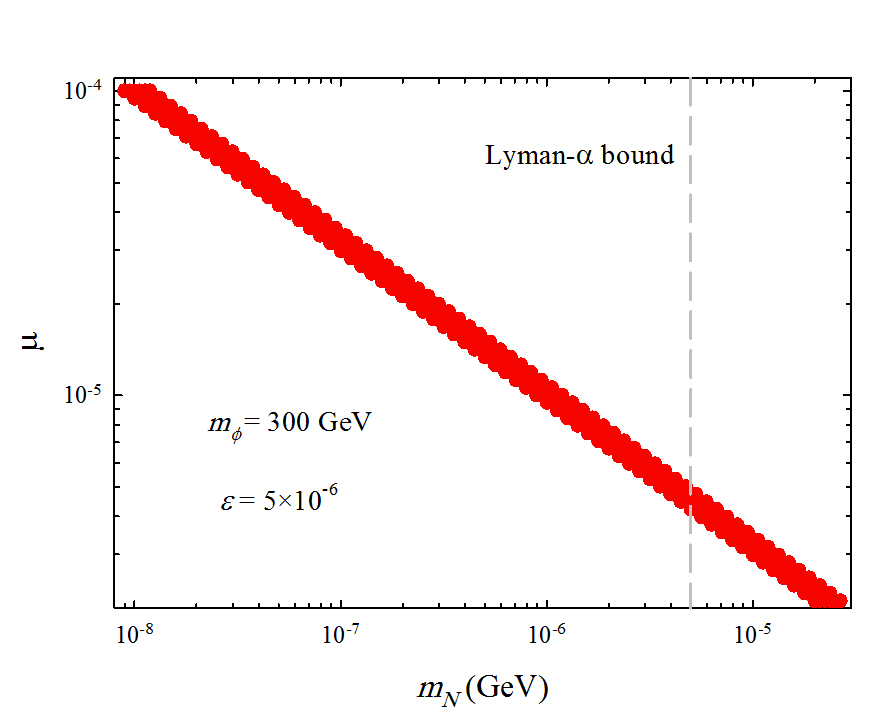}
	\caption{
		Viable parameter space in the $(m_N,\mu)$ plane.
		The red diagonal band corresponds to parameter combinations that reproduce the observed DM relic density while satisfying the freeze-in production condition.
		The vertical dashed line represents the Lyman-$\alpha$ lower bound on $m_N$.
		The allowed region corresponds to the overlap between the band and the	region to the right of the Lyman-$\alpha$ constraint.
	}
	\label{fig:mu_mN}
\end{figure}

\begin{figure}[!htbp]
	\centering
	\includegraphics[width=0.75\textwidth]{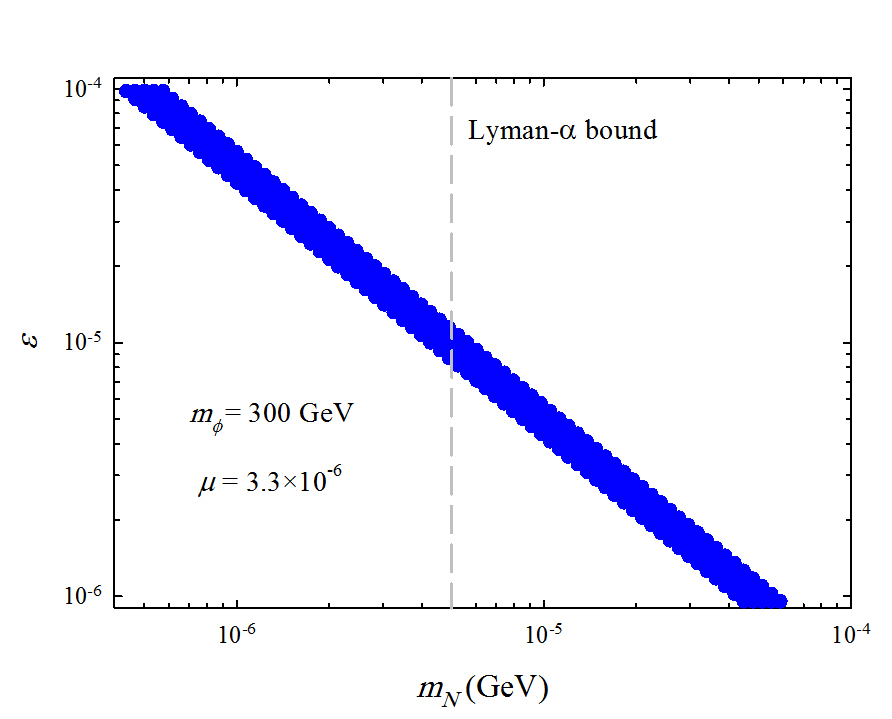}
	\caption{
		Viable parameter space in the $(m_N,\varepsilon)$ plane.
		The blue diagonal band corresponds to the correct relic density
		and consistent freeze-in production.
		The vertical dashed line represents the Lyman-$\alpha$ lower bound on the sterile neutrino mass.
		The allowed parameter space lies at the intersection of the band
		and the region to the right of the Lyman-$\alpha$ constraint.
	}
	\label{fig:eps_mN}
\end{figure}

Similarly, Fig.~\ref{fig:eps_mN} presents the viable parameter space in the $(m_N,\varepsilon)$ plane for fixed $m_\phi = 300\,\mathrm{GeV}$ and $\mu = 3.3 \times 10^{-6}$. The blue diagonal band corresponds to parameter combinations that reproduce the observed DM relic abundance while satisfying the freeze-in production condition. The inverse correlation between $m_N$ and $\varepsilon$ follows directly from the relic density requirement, implying that a larger sterile-neutrino mass must be compensated by a smaller CP asymmetry parameter. As in the previous case, the viable region is restricted to the parameter space to the right of the Lyman-$\alpha$ bound.

Overall, the model admits a well-defined region of parameter space where asymmetric sterile neutrino DM can account for the observed relic abundance while satisfying relevant cosmological and phenomenological bounds. The interplay between the mass of the sterile neutrino, the effective coupling $\mu$, and the CP-violating parameter $\varepsilon$ ensures a predictive and constrained framework for AFI DM production.

We emphasize that the viable regions shown in Figs.~\ref{fig:mu_mN} and~\ref{fig:eps_mN} are representative rather than exhaustive. Different choices of the model parameters would shift the allowed bands accordingly, while preserving the overall qualitative behavior. Future improvements in Lyman-$\alpha$ observations and small-scale structure measurements may further reduce the allowed parameter space and provide stronger tests of the model.

\section{Concluding remarks}\label{sec:conclud} 
In this work, we have investigated a minimal and predictive framework for asymmetric sterile neutrino DM produced via the freeze-in mechanism. The model extends the SM by introducing a scalar mediator $\phi$, a sterile neutrino $N$ as the DM candidate, and a singlet fermion $\chi$ participating in the decay process $\phi \rightarrow \bar{\chi} N$. The production of DM occurs out of thermal equilibrium through the decay of the mediator field, while a CP-violating parameter $\varepsilon$ generates the sterile-sector asymmetry and the trilinear coupling $\mu$ determines the efficiency of its production.

We derived and numerically solved the relevant Boltzmann equations to determine the evolution of the asymmetric yield. The final relic abundance is controlled by the interplay between the sterile neutrino mass $m_N$, the effective coupling $\mu$, and the CP-violating parameter $\varepsilon$. We showed that the present-day DM density, $\Omega_{\rm DM} h^2 \simeq 0.12$~\cite{Planck:2018}, can be naturally reproduced within well-defined regions of parameter space through freeze-in production, without thermalizing the sterile sector with the SM plasma.

The momentum distribution of sterile-neutrino DM is non-thermal, since the particles are produced through freeze-in decays of the scalar mediator. We demonstrated that the resulting average momentum satisfies $\langle p/T \rangle < 3.15$, indicating that the DM candidate is colder than thermal warm DM scenarios~\cite{Merle:2013,Merle:2015b}. Consequently, the model can be consistent with Lyman-$\alpha$ structure formation bounds~\cite{Palanque-Delabrouille:2019,Irsic:2024}, provided the sterile neutrino mass exceeds the corresponding lower limit derived in Sec.~\ref{sec4.3}.

We further imposed the freeze-in production condition ($\Gamma_\phi < H$)~\cite{Hall:2009} and the Lyman-$\alpha$ structure formation constraint~\cite{Palanque-Delabrouille:2019,Irsic:2024}. Combining these requirements with the observed DM relic abundance leads to well-defined viable regions in the $(m_N,\mu)$ and $(m_N,\varepsilon)$ parameter planes, shown in Figs.~\ref{fig:mu_mN} and~\ref{fig:eps_mN}. The resulting bands reflect the condition $m_N Y_-^\infty \simeq \mathrm{const}$ required to reproduce the observed DM density.

Overall, the scenario provides a simple and self-consistent realization of asymmetric sterile neutrino DM within a freeze-in framework. The model remains compatible with current cosmological and phenomenological observations while offering predictive correlations among its fundamental parameters. Future cosmological and astrophysical observations, including improved Lyman-$\alpha$ measurements and CMB-S4~\cite{Abazajian:2016}, may further test the AFI framework and probe the viability of asymmetric sterile-neutrino dark matter.

\section*{ACKNOWLEDGMENTS}
The author is grateful to Seyed Yaser Ayazi, Michael Schmidt, Emiliano Molinaro, and James Unwin for insightful discussions and valuable scientific guidance.





\end{document}